# Molecular Dynamics Studies of Changes in the DNA-Structure as Result of Interactions with Cisplatin

Alexander N. Issanin and Walter Langel

*Institut für Chemie und Biochemie, Universität Greifswald,*

*Soldmannstraße 23, 17489 Greifswald, Germany*

Patrick J. Bednarski

*Institut für Pharmazie, Universität Greifswald,*

*F.-L.-Jahnstraße 17, 17489 Greifswald, Germany*

Abstract

The 3'-monofunctional adduct of cisplatin and d(CTCTG*G*TCTC)$_2$ duplex DNA in solvent with explicit counter ions and water molecules were subjected to MD-simulation with AMBER force field on a nanosecond time scale. In order to simulate the closure of the bond between the Pt and 5'-guanine-N7 atoms, the forces acting between them were gradually increased during MD. After 500-800 ps the transformation of the mono-adduct (straight DNA with the cisplatin residue linked to one guanine-N7) to the bus-adduct (bent DNA where Pt atom is connected through the N7 atoms of neighboring guanines) was observed. A cavity between palatinate guanines is formed and filled with solvent molecules. The rapid inclination of the center base pairs initiates a slow transition of the whole molecule from the linear to the bent conformation. After about 1000-1300 ps a stable structure was reached, which is very similar to the one described experimentally. The attractive force between the Pt-atom and the N7 of the second guanine plays the main role in the large conformational changes induced by formation of the adduct-adduct. X-N-Pt-N-torsions accelerate the bending but a torsion force constant greater than 0.2 Kcal/mol lead to the breaking of the H-bonds within the base pairs. The present study is the first dynamical simulation that demonstrates in real time scale such a large conformational perturbation of DNA.

Introduction

Nearly all mechanisms of cisplatin cytotoxic activity proposed to date consider DNA to be the primary biological target [1]. Interactions with other cellular components have also been observed, but there is no convincing evidence that these are involved in the mechanism of drug action [2]. The major products of DNA-platination are 1,2-intrastrand cross-links involving adjacent bases with cis-[Pt(NH$_3$)$_2${d(GdG)}] (cis-GG) and cis-[Pt(NH$_3$)$_2${d(AdG)}] (cis-AG) comprising 47-50% and 23-28% of the adducts formed respectively. The remaining 10-13% consist of mono-functional derivatives of guanine and products of 1,3-intrastrand cross-platination [3].

By the mid 90's the kinetic pathways of the cisplatin-DNA interaction were elucidated [4]. In a hydrolysis reaction with a half-life of 2.3 h at 37 °C one chloride ligand in cisplatin is replaced by water [5]. The product of this partial hydrolysis cis-[Pt(NH$_3$)$_2$(H$_2$O)Cl]$^+$ reacts rapidly with DNA, forming a mono-adduct with either adenine or guanine. These mono-adducts were postulated to be inactive in processes leading to the death of cancer cells [3] because other platinum compounds which bind to DNA only in monofunctional manner show no anticancer activity.

Following the hydrolysis of the second Pt-Cl bond, the G-monoadduct undergoes a rapid chelation reaction and forms a bis-adduct if the 5'-base adjacent to the guanine is A or G. This reaction perturbs the global structure of the DNA-molecule with respect to B-DNA so that a significant curvature around the platination site is induced [6]. The structures of these bis-adducts were elucidated in crystal by X-ray diffraction [7,8] and in solution by NMR spectroscopy [9,10], respectively. Both results are noticeably different, presumably because the interaction between adjacent DNA molecules leads to significant deformations in the crystal. However, in both sets of data the H-bonds between cytosine and corresponding platinated guanine are conserved.

One of the factors thought to determine the anticancer activity of cisplatin is the geometry of the adduct with DNA [3]. A major role is played by the extension of the width of the minor groove during the bending process, by which a nonpolar part of DNA becomes accessible for solvent water molecules. In addition, the increased width allows specific interactions with proteins that would not otherwise bind with high affinity to normal B-DNA, such as HMG-group proteins [11-15].

Prediction of the final structure for adducts of DNA with platinum complexes other than cisplatin will only be possible when dynamics of the bending mechanism are understood in detail. To date, such studies are lacking. The linear structure of DNA is stabilized in solution by significant stacking stabilization energies and hydrophobic interactions, but still the energy gain due to the formation of only one additional Pt-N-bond must be sufficient to favor the bent structure of the complex with cisplatin. Three mechanisms of chelation are conceivable: (i) the entire DNA molecule is permanently bent to some extent as a result of Brownian motion or interactions with DNA-binding proteins. As soon as such a deformation leads to an appropriate position of Pt and N7-atom of nonplatinated guanine and allows for covalent bond formation, further irreversible bending of the whole DNA structure will occur; (ii) two neighboring base pairs, one platinated and the other not, have some degrees of freedom so that they could incline with respect to each other due to thermal motion and reach a suitable position to form the bis-adduct; (iii) the attractive forces between Pt(II) and 5'-N7 result in an inclination of two base pairs with respect to each other, following formation of a new bond. The aim of this study was to understand the details of this bending process.

3'G-monoadduct as initial point was chosen comparing 3D-structures of two possible monoadducts. Transformation of 5'-monoadduct to bis-adduct seems to be less probable because the distance between potential reaction centers (Pt and N7) in 5'-monoadduct is about 5.5Å, what is greater than one in 3'-monoadduct (3.6 Å). This assumption is also confirmed by the literature data [16].

AMBER is a standard force field used in DNA simulations. The versions up to 3.0 [17,18] treated the solvent as a continuum with adjustable dielectric constant [19]. As of AMBER 4.1 [20] solvent molecules can be taken into account explicitly. This was applied to DNA [21], RNA [22] and a copolymer [22]; in those simulations average helical parameters and their fluctuations were measured. With the AMBER force-field the canonical B-DNA structure in water solution can be well reproduced for several nanoseconds [23-26].

Geometry optimization has been done for platinum complexes with small ligands [27-29] such as two bases, and force field parameters for the bonds with Pt(II) have been derived. Molecular dynamics simulations of the cisplatin-DNA complex with ten base pairs were performed by Scheeff et al. [30] *in vacuo*. As electrostatic interactions are overestimated without water, the authors had to decrease the charge for phosphate groups and applied atomic constraints to tie the base pairs on both ends of oligomer in order to keep both DNA strands together. This simulation resulted in a stable trajectory for the bent structure. No realistic simulation of the bending process itself is possible, however, without taking into account the hydrodynamics of the solvent.

Computational Details

For molecular dynamic (MD) simulation the AMBER 5.0 software package [31] has been used which includes version 4.1 of the AMBER force field [20]. The AMBER 3.0 force field has been extended to Pt(II) compounds in [32]. We transferred these additional parameters to AMBER 4.1 without modification. The partial charges for the atoms in the guanines are significantly modified by the platination due to electron withdrawal to the positively charged platinum atom (tab 1). We applied the difference between platinated and free guanines described in [32] to the atomic charges from AMBER 4.1. A similar charge transfer technique has been used before in force field development [33,34].

We generated a DNA d(CTCTGGTCTC)$_2$ duplex using the NUCGEN script included in the AMBER 5.0 package. The length was limited to 10 base pairs rather than 12 in the experiments in order to speed up calculations. The mono-adduct of cisplatin and DNA was created by connecting a Pt(NH$_3$)$_2$ residue to the N7 atom of the 3´ guanine, the Pt lying in the guanine plane. All parameters for this Pt-N7 bond were set to their normal values. The resulting distance between Pt and N7 in the second guanine was about 3.6 Å, which is almost twice the bond length. The cisplatin-DNA complex was surrounded with a solvent bath, having rectangular shape (38x41x51 Å) and consisting of approximately 3500 TIP3P [35] water molecules. Each strand of the duplex contains 9 negative charges located on the phosphate groups and the cisplatin residue has an electric charge of +2. 16 Na$^+$ ions were added providing electroneutrality of the entire system. They were placed around the solute molecule calculating a Coulomb potential grid with LEaP program in the AMBER package. This procedure finds appropriate positions for Na$^+$ ions by minimizing the electrostatic energy of the system. The solvent box provides 8 Å minimum distance between DNA-atoms and the walls of the box in Y- and Z- and 10 Å in X-direction allowing some additional space for bending of DNA chain in the YZ-plane. According to the AMBER MD protocol, the rotational and translational motions of the solute molecule are regularly removed and the appropriate position of the DNA-adduct in the anisotropic box is provided during the entire MD-trajectory.

The equilibration of the system is described elsewhere [25] in detail. MD simulations were performed in the NTP-ensemble with a 2 fs time step. During the equilibration stage the volume was adjusted so that the pressure in the system was approximately 1 atm. The Berendsen algorithm for temperature bath coupling [36] was utilized to keep the temperature of the core reasonable close to that of solvent.

**Table 1.** Atomic charges

| Atom | Free base charges | Platinated charges |
|---|---|---|
| N1 | -0.5053 | -0.5053* |
| H1 | 0.3520 | 0.3520* |
| C2 | 0.7432 | 0.7432* |
| N2 | -0.9230 | -0.9230* |
| H21, H22 | 0.4235 | 0.4235* |
| N3 | -0.6636 | -0.6636* |
| C4 | 0.1814 | 0.2044 |
| C5 | 0.1991 | 0.2681 |
| C6 | 0.4918 | 0.5148 |
| O6 | -0.5699 | -0.5699 |
| N7 | -0.5725 | -0.3435 |
| C8 | 0.0736 | 0.2466 |
| H8 | 0.1997 | 0.1187 |
| N9 | 0.0577 | 0.0807 |
| cisplatin residue | | |
| Pt | | 0.7090 |
| N | | -0.6480 |
| H | | 0.2780 |

* Original AMBER charges.

During the production stage, atomic coordinates of the solute molecule were stored every 200 fs. All bond lengths were fixed by SHAKE-constraints [37] with exception of the Pt-N bond, which was forming during the bending process.

Trajectory files were analyzed by using Cerius-2 4.0 and Insight-II 98.0 [38]. Helical parameters of the DNA-duplex were obtained from Curves 5.3 [39,40]. According to this algorithm, total bending or global curvature (UU) is calculated as the angle between the orthogonal projection of the base planes in 2nd and n-1st base pairs of the nucleotide. Local parameters such as Rise, Tilt, Shift and Roll characterize the mutual arrangement of neighboring base pairs along Z-axis. Since both local and global parameters are subjected to intense thermal fluctuations, floating averages of the data have been plotted.

In the AMBER force field hydrogen bonds are modeled by a combination of appropriate electrostatic and vdW interactions between acceptor and donor atoms. During a molecular rearrangement, process such as the simulated H-bonds can be broken. Their conservation in base pairs with platinated guanines was checked by monitoring the local helical parameters Shear, Stretch, and Stagger, which are zero in canonical B-DNA.

**Table 2.** Local and global parameters for cisplatin-DNA complex

| Parameter | X-ray | NMR | MD(X-ray) | MD(NMR) | A* | B* | C* |
|---|---|---|---|---|---|---|---|
| Shift G5-G6 (Å) | 1.66 | 0.46 | -0.07±0.74 | -0.58±0.63 | -0.12±0.72 | 0.45±0.47 | 0.54±0.48 |
| Slide G5-G6 (Å) | -1.10 | -0.39 | -0.78±0.52 | -0.90±0.54 | -1.9±0.40 | -0.55±0.33 | -0.38±0.34 |
| Rise G5-G6 (Å) | 2.23 | 4.34 | 4.37±1.2 | 5.11±0.85 | 3.9±0.80 | 4.8±0.65 | 4.8±0.61 |
| Tilt G5-G6 (deg) | 1.94 | -4.33 | 3.7±4.9 | -2.3±5.2 | -9.8±5.0 | -2.3±4.0 | -1.8±4.0 |
| Roll G5-G6 (deg) | 23.48 | 45.83 | 52.3±7.7 | 54.8±5.4 | 35.3±6.3 | 49.0±6.2 | 47.3±6.3 |
| Twist G5-G6 (deg) | 26.74 | 3.04 | 13.4±13 | 7.4±7.4 | 21.4±4.4 | 13.7±5.8 | 12.8±5.9 |
| Minor groove width** (Å) | 10.45 | 10.74 | 11.4±1.1 | 11.7±1.4 | 7.9±1.6 | 11.27±0.93 | 11.69±0.74 |
| H-bonds*** G5-C16 | + | + | + | + | - | + | + |
| H-bonds*** G6-C15 | + | + | + | + | + | + | + |
| Overall bending (deg) | 39.0 | 78.3 | 57±16 | 68±12 | 59±13 | 55±12 | 47±11 |
| Shortening (%) | 4.77 | 20.45 | 13.2±5.2 | 16.1±3.7 | 12.0±4.1 | 9.6±3.1 | 7.4±2.7 |

*- averaging is started 200 ps after jump (750 ps, 900 ps and 1190 ps consequently).

** - in G*G* region.

*** - based on distances and angles analysis.

Results and discussion

In order to check the force field two 1 ns MD trajectories of the adduct consisting of 10 DNA base pairs and one cisplatin residue have been simulated, starting with NMR [10] and X-ray [7] structures, respectively. The stability and convergence behavior of the simulations was monitored by two-dimensional root mean square deviation (rms) maps. In the beginning of each run the structure of the bis-adduct drifts away form the initial structure. During the second part of each trajectory, the observed rms deviations resulted from thermal motion of one basic conformation rather than drift or fluctuation among two or more structures, showing that the equilibrium state was reached. This suggests that very complete relaxation of the initial structure occurs on the time scale of our simulation, and that molecular dynamics can be used for geometry refinement of the system.

In each run the system rapidly reached local equilibrium for platinated base pairs after 10-50 ps, whereas the rearrangement of the bulky rest of the molecule was much slower. Local and global helical parameters (tab. 2) show that both MD results are very similar to the NMR-structure. Consequently convergence was fast for the NMR-run, taking 100-200 ps, whereas the trajectory starting from X-ray data reached the final structure only after 300-400 ps. This behavior is to be expected because the MD-simulation cell represents the solution rather than the crystal.

The simulated structure shows a slightly higher local deformation of the platinated guanine base pairs, as indicated by the Rise and Roll values, and a slightly smaller overall bending than the experimentally determined structure in solution. The experimental values are, however, very close to or within the range of thermal fluctuation of the simulation results (tab. 2).

Having validated the force field for the DNA-cisplatin complex, we started the simulation of the bending process. These experiments were not intended to model the reaction coordinate for the substitution at the Pt(II) ion, which is widely believed to occur by way of a five-coordinated transition state. Rather, we were primarily interested in understanding how the formation of a single N-Pt-bond can dramatically disrupt the structure of B-DNA. A first minimization and molecular dynamics run without rescaling of the Pt-N stretching constant [32] resulted in DNA structures, where the reaction center was completely deformed because the two guanines attain a 90° angle already during the minimization and the first MD steps. The rest of the molecule could not relax in such a short period of time.

As far as the inclination of the two guanines is induced by a reduction of the length of the newly forming Pt-N7 bond, this results only in an increase of the Tilt-angle between the two bases. By including significant torsional forces around the C-N bonds adjacent to the Pt, Roll increases, as well, which weakens the hydrogen bonds between the guanines and their cysteine partners. As a consequence, these bonds break, and the cysteines do not follow the inclination of the guanines. The rupture was indicated by a an increase of Shear and Stretch to about 0.5 - 1 Å, Stagger attained a value of -2 Å, while the initial value of those three variables is about zero. During several hundreds of picosecond this bis-adduct shows no tendency to recover from this non-appropriate conformation and fails to attain a structure that better corresponds to the experimental data.

**Table 3.** Rescaling of different force constants

| type | qty. | value[32] | traj. | **Time (ps): 0-120** | 120-220 | 220-320 | 320-420 | 420-520 | 520-rest |
|---|---|---|---|---|---|---|---|---|---|
| bond | | kcal/(mol·Å$^2$) | | | | | | | |
| Pt-NB2 | 1 | 366 | A | 0 | 3.66 | 36.6 | 366 | 366 | 366 |
| | | | B | 0 | 3.66 | 36.6 | 366 | 366 | 366 |
| | | | C | 0 | 3.66 | 36.6 | 366 | 366 | 366 |
| | | kcal/(mol·rad$^2$) | | | | | | | |
| NB2-PT-X | 3 | 42 | A | 0 | 0.42 | 4.2 | 42 | 42 | 42 |
| | | | B | 0 | 0.42 | 4.2 | 42 | 42 | 42 |
| | | | C | 0 | 0.42 | 4.2 | 42 | 42 | 42 |
| X-NB2-PT | 2 | 20 | A | 0 | 0.2 | 2.0 | 20 | 20 | 20 |
| | | | B | 0 | 0.2 | 2.0 | 20 | 20 | 20 |
| | | | C | 0 | 0.2 | 2.0 | 20 | 20 | 20 |
| | | kcal/mol | | | | | | | |
| X-NB2-PT-X | 6 | 0.5 | A | 0 | 0.005 | 0.05 | 0.05 | 0.1 | 0.2 |
| | | | B | 0 | 0.005 | 0.05 | 0.05 | 0.1 | 0.1 |
| | | | C | 0 | 0.005 | 0.05 | 0 | 0 | 0 |
| X-NB1-PT-NB2 | 2 | 0.5 | A | 0 | 0.005 | 0.05 | 0.05 | 0.1 | 0.2 |
| | | | B | 0 | 0.005 | 0.05 | 0.05 | 0.1 | 0.1 |
| | | | C | 0 | 0.005 | 0.05 | 0 | 0 | 0 |
| X-CB-NB2-PT | 2 | 2.55 | A | 0 | 0.0255 | 0.255 | 0.255 | 0.51 | 1.02 |
| | | | B | 0 | 0.0255 | 0.255 | 0.255 | 0.51 | 0.51 |
| | | | C | 0 | 0.0255 | 0.255 | 0.255 | 0 | 0 |
| X-CK-NB2-PT | 2 | 10 | A | 0 | 0.1 | 1 | 1 | 2 | 4 |
| | | | B | 0 | 0.1 | 1 | 1 | 2 | 2 |
| | | | C | 0 | 0.1 | 1 | 0 | 0 | 0 |

CK, CB - standard AMBER atom types

PT, NB1, NB2 - atom types from [32]

X - any atom

The harmonic model does not guarantee that the total distortion energy is always less than the Pt-N binding energy for highly stretched bonds such as the Pt-N bond with an initial length of 3.6 Å. A realistic model for two atoms taking part in the formation of chemical bond would be a Morse potential, which is, however, not provided by the AMBER force field. We, therefore, modeled the closure of the bond between Pt and N atom with the standard harmonic potential by varying the force constant. Gradual increasing of the stretch constant must provide smooth closure of reacting centers. Only at the end of this process will the distance and force constant between two atoms correspond to normal values as defined by molecular structure [41]. This approach provides that the force acting between the Pt and the N7 is smaller at a longer bond distance, which is an essential feature of a realistic interatomic pair potentials, and thus

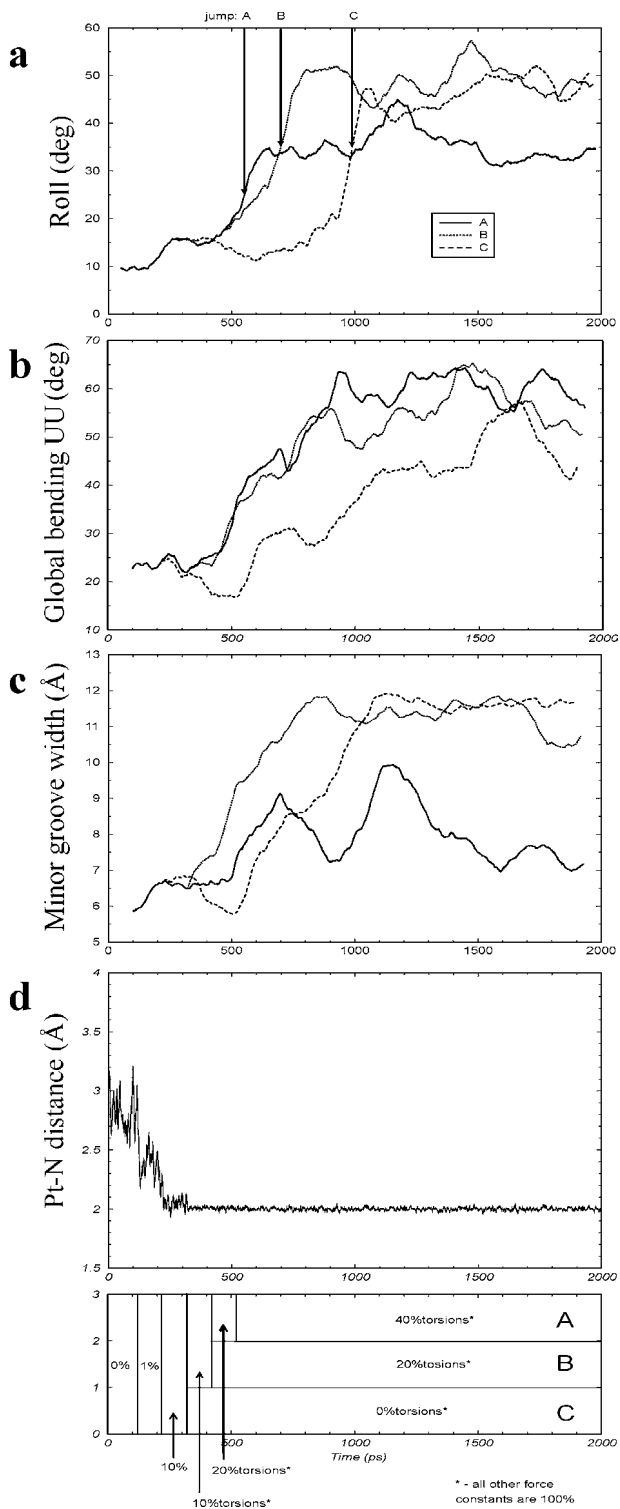

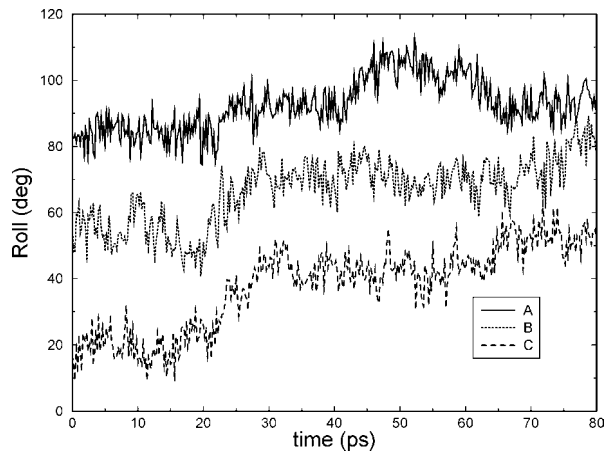

**Fig 2:** Time dependence of Roll during the jump. Traces **A** and **B** were shifted by 60° and 30° respectively. t = 0 corresponds to simulation times of 560 ps (**A**), 680 ps (**B**) and 960 ps (**C**) respectively. A sharp increase for **B** and **C** is seen after t = 20 ps during next 10 ps.

**Fig 1:** Results of the simulations of the bending process of DNA as induced by platination. Geometry parameters as evaluated from CURVES are plotted as a function of time. For clarity floating averages were calculated with appropriate time windows.

a      Roll over time (averaged over 100ps)

b      global bending (UU) (averaged over 200ps)

c      minor groove width (averaged over 200ps)

d      Pt-N distance (averaged over 1ps)

prevents an overestimation of the bond energy. To check this we traced the deformation energy:

$$E = E_b + E_a + E_t = \sum \frac{K_b}{2}(r - r_0)^2 + \sum \frac{K_a}{2}(\theta - \theta_0)^2 + \sum \frac{K_t}{2}[1 + \cos(n\varphi - \varphi_0)]$$

where $K_b$, $K_a$ and $K_t$ are stretching, bending and torsion force constants respectively; r, $\theta$, $\varphi$, $r_0$, $\theta_0$ and $\varphi_0$ are current and equilibrium distances, angles and torsions which include the Pt-N7(2) pairs. We compared the deformation energy as a function of the Pt-N7 bond length with the energy gain due to the closure of the Pt-N- bond. This has been calculated by density functional theory [42] in terms of a Pt-N-pair potential, which is fitted by a Morse potential with bond length, dissociation energy, and vibrational frequency of 1.7 Å, 86 kcal/mol, and 1139 cm$^{-1}$, respectively. For all distances attained during the MD-run the deformation energy is smaller than the respective energy gain due to the Pt-N attractive interaction.

The detailed protocol of the simulation with a stepwise increase of the force constants was as follows: During the first 100 ps after equilibration of the solvent bath all energy terms for Pt-N7 with the exception of nonbonded interactions are set to zero. Thus the bond between Pt and the N7 of the second guanine does not appear to exist, and the system behaves as mono-adduct (fig 1a). For the next two intervals of 100 ps each, 1% and 10% of the full values for these force constants, respectively, were applied (tab 3). After 300 ps stretching and bending, force constants were set to their standard values. As the four force constants for the torsions, which include Pt and N7, are essential for the kinetics of the bending process, three continuation runs with different rescaling were then started. During the first run (denoted **A**), the torsion force constants were reduced to 10% and 20% of the full value for the first and second interval of 100 ps, respectively, and to 40% for another 1500 ps. Run **B** consisted of 100 ps of 10% torsions followed by 1600 ps of 20%, and in **C** torsion force constants were set to zero for 1700 ps (see table 3 and fig 1).

The Pt-N7 distance is gradually decreasing from the start and attains its nominal value after 300 ps. Switching on the full stretching force constant then significantly reduces the thermal fluctuation of this parameter (fig 1a). As a consequence of this decrease, the guanine-guanine angle (Roll) and the minor groove width increase to about 15 ° and 6.8 Å, which is, however, far from the final values for the bent conformation (figs 1b and c).

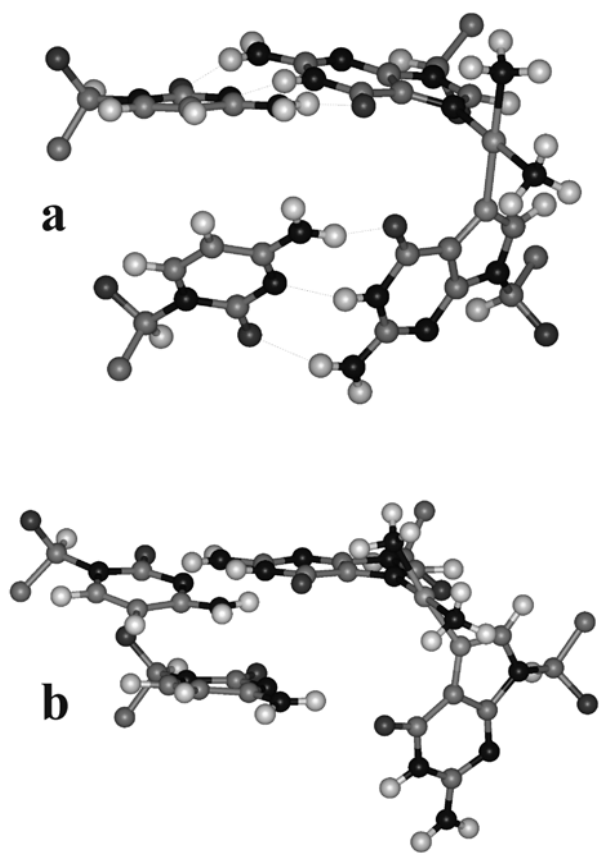

**Fig 3:** Structure of bis-adduct with conserved (**a**) and broken (**b**) H-bonds.

This slightly bent structure was only metastable, and in all three runs a very sharp rise of the angle between platinated guanines (Roll) from 15° to about 35° for **A** and 50° for **B** and **C**, respectively, was observed. Thereby the local conformation of Pt and two guanines flipped from a strained into a relaxed structure. With vanishing torsion, i.e. only with a stretching force constant this started after 980 ps (**C**), and by increasing the torsion force constant the initiation time was reduced to 700 and 580 ps for **B** and **A** respectively.

The inclination process of the base pairs took less than 10 ps (fig 1a, 2) and was accompanied by an increase in their mutual distance (Rise) by about 2 Å. Twist is hardly affected by the closing of the bond, whereas Tilt is reduced to about 14°. The thermal fluctuations of these local parameters are higher in the bis-adduct than in the mono-adduct, indicating significantly reduced stacking interactions within the deformed DNA duplex. The H-bonds of the platinated base pairs were stable in cases **B** and **C** while during trajectory **A** they were broken in one base pair exactly at the time of the flip, very similar to the calculation without reducing the torsional force constants at all (fig 3). Observed changes of Shear, Stretch and Stagger were very similar to those in the run without rescaling (see above). This was in contrast to **B** and **C**, where all those variables remained about zero. The resulting transformations were irreversible on the time scale of the simulation because the MD run of another 1 ns did not result in appropriate structure.

Fig 1d shows that our simulations result in a significant overall bending of the DNA-chain. For nonplatinated B-DNA, the average value of the bending depends on nucleotide sequence and length of oligonucleotide and fluctuates around zero with an amplitude of about 20°. The overall shape of the equilibrated DNA duplex was checked by averaging over the structures starting 200 ps after the "jump" (table 2). In contrast to the local deformation around the platination center, the global structure did not depend on the torsion force constant, and the bending is similar after all three runs. We consider the bending as conformational rearrangement that is initiated in one point and then spreads over the rest of the entire oligomer. Stacking and hydrophobic interactions, which normally stabilize the linear structure of B- DNA, no longer balance the energy gain by relaxing the platination center.

During the bending process, the width of minor groove between the two platinated base pairs is increasing (fig 1c) while the corresponding values determined for the ends of oligonucleotide remain without significant changes. Experimentally it was observed that the minor groove widens from 5.5Å for normal B-DNA to 10.8 Å (X-ray) or 11.2 Å (NMR). Analysis of water distribution around the target molecule revealed that the cavity between platinated guanines forming after "jump" does not remained empty. As B-DNA has no water molecules between base pairs, the alteration of the native structure caused by cisplatin affects solvent accessibility of the molecule. Run **B** showed that such a cavity is stable for at least 1000 ps after its formation.

Conclusion

Previous work in the field was extended by simulating the entire bending process rather than only initial and final states. This was possible by including molecular dynamics into the simulation using AMBER4.1 and enhancing it by parameters for the cisplatin-DNA adduct. In the experimental bent structures, the hydrogen bonds between all base pairs are conserved. The simulation shows that these weak bonds must be conserved during the whole bending process, as they will not form again afterwards. This is a delicate test for the force field. In the frame of AMBER, the closure of the bond was possible by gradually increasing the force constants for this additional bond from zero for the mono-adduct to the full values of the bis-adduct. In particular, the X-Pt-N7-X torsions tend to impose stress on the hydrogen bonds and must be reduced well below their standard values at the beginning of the simulation.

From the simulation a detailed mechanism can now be derived which answers the question as to how a single additional bond can result in the global deformation of a large molecule with a stable linear structure. MD simulation on DNA showed that Brownian motions neither of the whole chain nor of the guanines in the chain can compete with the attractive interaction between the Pt and the N7 atoms for the chelation process, and we thus favor model (iii) discussed in the introduction.

Our mechanism provides evidence for two different relaxation processes, an intramoelcular flipping of the base pairs in the 100 GHz range and a much slower global relaxation resulting in an overall bending. It is in the range of 1-10 GHz, but may be much slower in larger DNA-strands due to hydrodynamic hindrance. Recently the dynamics of DNA in solution has been studied by inelastic neutron scattering [43], and the data have been interrupted in terms of two processes, on peaking around 1000 GHz, the other one being slower than about 10 GHz. It is not possible to assign these maxima directly to our simulations, but the experiment confirms that a distinction of very different time scales for the relevant relaxation processes is feasible.

Acknowledgment

We gratefully acknowledge financial support by the Fond der Chemischen Industrie.